\documentclass[sigconf]{acmart}
\usepackage{multirow}
\usepackage{graphicx}
\usepackage[tight,footnotesize]{subfigure}

\copyrightyear{2024}
\acmYear{2024}
\setcopyright{acmlicensed}
\acmDOI{10.1145/3627673.3679916}

\acmConference[CIKM '24] {Proceedings of the 33rd ACM International Conference on Information and Knowledge Management}{October 21--25, 2024}{Boise, ID, USA}

\acmISBN{979-8-4007-0436-9/24/10}

\begin{document}

\title{Enhancing SPARQL Generation by Triplet-order-sensitive Pre-training}
\author{Chang Su}
\orcid{1234-5678-9012}
\affiliation{%
  \institution{Shanghai Jiao Tong University}
  \city{Shanghai}
  \country{China}
}
\email{suchang0912@sjtu.edu.cn}

\author{Jiexing Qi}
\affiliation{%
  \institution{Shanghai Jiao Tong University}
   \city{Shanghai}
    \country{China}
}
\email{qi_jiexing@sjtu.edu.cn}

\author{He Yan}
\affiliation{%
  \institution{ProtagoLabs Inc.}
  \city{Vienna}
  \state{Virginia}
    \country{USA}
}
 \email{he.yan@protagolabs.com}

\author{Kai Zou}
\affiliation{%
  \institution{NetMind.AI LTD}
  \city{London}
    \country{UK}
}
\email{kz@netmind.ai}

\author{Zhouhan Lin}
\authornote{Corresponding Author.}
\affiliation{%
  \institution{Shanghai Jiao Tong University}
   \city{Shanghai}
    \country{China}
}
\email{lin.zhouhan@gmail.com}

\renewcommand{\shortauthors}{Chang Su, Jiexing Qi, He Yan, Kai Zou, and Zhouhan Lin}

\begin{abstract}
Semantic parsing that translates natural language queries to SPARQL is of great importance for Knowledge Graph Question Answering (KGQA) systems. Although pre-trained language models like T5 have achieved significant success in the Text-to-SPARQL task, their generated outputs still exhibit notable errors specific to the SPARQL language, such as triplet flips. To address this challenge and further improve the performance, we propose an additional pre-training stage with a new objective, Triplet Order Correction (TOC), along with the commonly used Masked Language Modeling (MLM), to collectively enhance the model's sensitivity to triplet order and SPARQL syntax. Our method achieves state-of-the-art performances on three widely-used benchmarks.\footnote{Our implementation is available at \url{https://github.com/LUMIA-Group/TosT5}.}
\end{abstract}




\maketitle

\section{Introduction}
\begin{figure*}
\centering
\includegraphics[width=\textwidth]{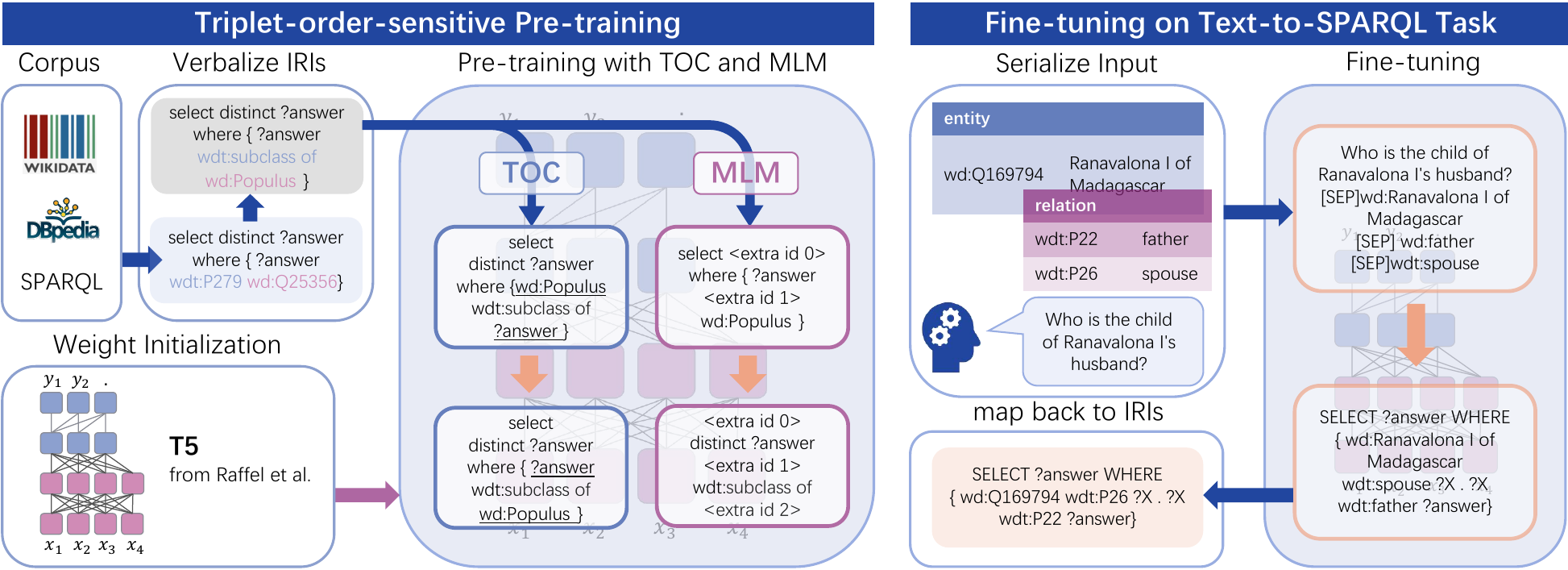}
\caption{The overview of our approach. The TosT5 model first undergoes the triplet-order-sensitive pre-training stage and then is fine-tuned on the downstream task.}

\label{fig:pipeline}
\end{figure*}
As the Semantic Web evolves rapidly, various Knowledge Graphs (KGs) have emerged, holding crucial significance in diverse domains, including KG-based question answering\cite{liu2018entity}, information retrieval\cite{liu2018entity} and recommendation systems\cite{guo2020survey}. Given a question posed in natural language, the Knowledge Graph Question Answering (KGQA) system's objective is to retrieve the correct answer from the KG\cite{rony2022tree}, thus empowering users, especially those lacking programming expertise, to effortlessly interact with KGs. The primary challenge is to convert natural language queries from users into a specialized query language for KGs, such as SPARQL.

A standard KGQA system typically involves three main steps\cite{banerjee2022modern}: 1) Entity Linking (EL), 2) Relation Linking (RL), and 3) Query Building (QB). Once the entities and relations are linked, the query-building module integrates this information into a formal SPARQL query. 
While EL and RL have become classical problems with substantial existing research\cite{mihindukulasooriya2020leveraging}\cite{lin2020kbpearl}\cite{dubey2018earl}, QB represents a unique challenge with fewer focused studies at present. However, its impact on the final performance of KGQA systems is significant. Thus, our paper focuses on QB, that is, generating SPARQL queries from natural language questions with gold entities and relations available.

Apart from early attempts using rule-based approach\cite{lan2022complex} which cannot scale to the diverse expressions found in natural language, more recent research on the Text-to-SPARQL task utilizes advanced pre-trained encoder-decoder models like T5\cite{raffel2020exploring} and BART\cite{lewis2019bart}, significantly outperforming traditional approaches. 
To further enhance these models' performance and reduce errors in generated queries, ReTraCk\cite{chen2021retrack} leverages a grammar-based decoder, and another work \cite{cao2021program} employs a function-based decoder, both applying schema-level constraints to prune the search space during decoding. 
These methods help to improve the syntax correctness. However, there are some errors, such as triplet-flip errors
, that do not compromise the grammatical correctness and executability of the queries and cannot be resolved through schema-level constraints.

In this work, we propose a \textbf{T}riplet-\textbf{o}rder-\textbf{s}ensitive \textbf{T5} (\textbf{TosT5}) model, which exploits a novel pre-training stage positioned between the general T5 pre-training and the task-specific fine-tuning for Text-to-SPARQL task, to address the triplet-flip errors, thereby enhancing pre-trained models for SPARQL generation. In pre-training, we introduce a novel objective called Triplet Order Correction (TOC), which helps TosT5 to better understand the sequential order of elements within triplets. This objective is combined with the commonly used Masked Language Modeling (MLM) objective, further enhancing the model's comprehension of the SPARQL language. To improve the semantic information contained in SPARQL queries and leverage the pre-trained model's ability in language understanding, we also propose to verbalize Internationalized Resource Identifiers (IRIs) in queries into their literal values during training. 
After undergoing the triplet-order-sensitive pre-training, the TosT5 is fine-tuned on the downstream Text-to-SPARQL task. 

Experiment results show that the TosT5 model significantly improves SPARQL query generation quality, achieving new state-of-the-art performances on three well-known KGQA datasets: LC-QuAD 2.0\cite{dubey2019lc}, QALD-9\cite{ngomo20189th}, and QALD-10\cite{usbeck2023qald}. Meanwhile, detailed error analysis and ablation studies validate the effectiveness of our pre-training objectives and verbalizing IRIs.

\section{Preliminaries}
\label{sec:preliminaries}
\textbf{Problem Setup.} Knowledge Graphs (KGs) store structured multi-relational data in the form of (subject entity, relation, object entity)\cite{dai2018learning}, where each of the three elements is identified by an Internationalized Resource Identifier (IRI). A KG can be represented as: 
\begin{center}
$\mathcal{KG}=\left\{(e_{s}, r, e_{o})\mid e_{s}, e_{o} \in \mathcal{E}, r \in \mathcal{R}\right\}$
\end{center}
where $\mathcal{E}$ and $\mathcal{R}$ represent the sets of entities and relations, respectively. A triplet can be written as $(e_{s}, r, e_{o})$, which consists of a subject entity $e_{s}$, a relation $r$, and an object entity $e_{o}$. 

The query-building module's objective within Knowledge Graph Question Answering (KGQA) system is to map a given natural language question $Q$ to a SPARQL query $Y$, using the provided gold entities $E = \{E_i\}_{i=1}^{|E|}$, where $E \subseteq \mathcal{E}$, and gold relations $R = \{R_i\}_{i=1}^{|R|}$, where $R \subseteq \mathcal{R}$. $E_i$ and $R_i$ are predefined IRIs in the KG, and their corresponding literal values are denoted as $L_{E_i}$ and $L_{R_i}$. For example, entity $E_1$ is \texttt{wd:Q25356}, with its literal value $L_{E_1}$ equal to \texttt{Populus}.

\noindent \textbf{Triplet-Flip Error.} 
Our analysis finds that\footnote{Error analysis in Section \ref{sec:errors} reveals the specific number of triplet-flip errors, which even account for up to 50\% of all prediction errors on the LC-QuAD 2.0 dataset.}, a large portion of errors in model's prediction comes from mistakenly predicting the order of elements within the triplets. We call this kind of error as \emph{triplet-flip error}. It is worth noting that flipping the elements in the triplet may not always cause grammar errors. If the positions of $e_{s}$ and $e_{o}$ are erroneously swapped, it will result in a grammatically correct SPARQL query, but yielding incorrect or unavailable results. As an example, let the target SPARQL query be

\begin{center}
\small
\texttt{select distinct ?ans where \{ ?ans wdt:P279 wd:Q25356\}}
\end{center}
where \texttt{wdt:P279} and \texttt{wd:Q25356} are the IRIs for relation \texttt{subclass of} and entity \texttt{Populus}. The model may predict 
\begin{center}
\small
\texttt{select distinct ?ans where \{ wd:Q25356 wdt:P279 ?ans\}}
\end{center}
instead, where the positions of $e_s$ and $e_o$ are mistakenly flipped. 

However, real-world scenarios are more complex than this simple example, where triplet-flip errors may occur in nested queries, queries with multiple triplets, etc., adding to the challenge.

\section{METHOD}

In this section, we introduce our TosT5 model, which chooses T5 as the backbone and exploits an additional novel pre-training stage specifically designed to address the triplet-flip errors, to enhance the SPARQL generation. In our approach, the pre-training uses two objectives to enhance the model's understanding of the SPARQL language. Verbalizing IRIs is proposed to help the model better understand. Figure \ref{fig:pipeline} shows an overview of our approach.
\subsection{Triplet-order-sensitive Pre-training}
\label{sec:pretraining}
We introduce a triplet-order-sensitive pre-training stage inserted between the general T5 pre-training and the fine-tuning on the Text-to-SPARQL task. During this stage, the TosT5 model is self-supervised by two objectives, namely Triplet Order Correction and Masked Language Modeling, to enhance the comprehension of task-specific knowledge. Accurate SPARQL queries are employed as the pre-training corpus.

\noindent\textbf{Triplet Order Correction.}
Given a SPARQL query, the Triplet Order Correction (TOC) objective focuses on grasping the correct order of elements within triplets, which is crucial for alleviating the triplet-flip errors that occurred in the prediction. To achieve it, we first randomly rearrange the positions of subject entity, relation, and object entity within each triplet in SPARQL queries. The original triplet $(e_{s}, r, e_{o})$ is transformed into a permutation, such as $(e_{o}, r, e_{s})$ or any other possible arrangement of $e_{s}$, $r$, and $e_{o}$ with equal probability. There is also a chance that the order remains unchanged. When dealing with multiple triplets, the shuffling of each triplet is independent of the others. 
For elliptical constructions, such as a semicolon indicating the omission of the subject entity, we restore the elliptical triplet back to its plain form $(e_{s}, r, e_{o})$ before processing. Finally, we expect the model to predict the initial order of element in triplets and reconstruct the original accurate SPARQL statements. Formally, given a SPARQL statement $x$, we transform it into $\tilde{x}$ as the model's input and expect the model to generate the original $x$, and the TOC loss is given by:
\begin{center}
$\mathcal{L}_{TOC}=-\sum_{i=1}^{|x|} \log P_{\Theta}\left(x_{i} \mid x_{<i} ; \tilde{x}\right)$  
\end{center}
where $\Theta$ denotes the model parameters.

\noindent\textbf{Masked Language Modeling.}
The Masked Language Modeling (MLM) 
aims to better understand the expression of the SPARQL language, 
randomly masking tokens in the input SPARQL statements, and then let the model predict them based on the context. 
We utilize a span masking strategy, aligning with the
general pre-training for the T5 model.
Spans of the input sequence $x$ are masked by sentinel tokens, and the output sequence $y$ is formed as a concatenation of the same sentinel tokens and the real masked tokens.
Similarly, the training
loss is denoted as:
\begin{center}
$\mathcal{L}_{MLM}=-\sum_{i=1}^{|y|} \log P_{\Theta}\left(y_{i} \mid y_{<i} ; x\right)$    
\end{center}
where $\Theta$ denotes the model parameters.

The triplet-order-sensitive pre-training stage combines the TOC and MLM in a multi-task learning framework, and the final loss is the sum of $\mathcal{L}_{TOC}$ and $\mathcal{L}_{MLM}$. 

\subsection{Fine-tuning on Downstream Task}
Following the pre-training, our TosT5 model is fine-tuned on the Text-to-SPARQL task, converting users' natural language question $Q$ into SPARQL query $Y$ with both gold entities and gold relations provided. Formally, the input $X$ is organized as:
$$X=\overline{Q|E_1\ L_{E_1}, E_2\ L_{E_2},\cdots|R_1\ L_{R_1},R_2\ L_{R_2},\cdots.}$$
and the training loss is defined by
\begin{center}
$\mathcal{L}_{ft}=-\sum_{i=1}^{|Y|} \log P_{\Theta}\left(Y_{i} \mid Y_{<i} ; X\right).$    
\end{center}


\subsection{Verbalizing IRIs}
\label{sec:transformation}
As mentioned in Section \ref{sec:preliminaries}, IRIs are employed to represent entities and relations in SPARQL queries. However, they merely serve as identifiers and lack semantic information, presenting as challenges to the model's comprehension.
Therefore, we propose to verbalize IRIs to literal values before the pre-training, to better leverage the pre-trained model’s ability in language understanding. And during fine-tuning, the literal values in the model's predicted outputs are mapped back to IRIs, to ensure the executability of the results. For example, we use \texttt{wd:Populus} in the inputs and outputs of the model rather than \texttt{wd:Q25356} during the pre-training. Finally, when the output is predicted, the literal value \texttt{wdt:spouse} is mapped back to the corresponding IRI \texttt{wdt:P26} for direct execution.

\section{EXPERIMENT} 

\subsection{Basic settings}
We conduct experiments on three widely-used question answering datasets based on Wikidata: LC-QuAD 2.0\cite{dubey2019lc}, QALD-9\cite{ngomo20189th} and QALD-10\cite{usbeck2023qald}. LC-QuAD 2.0 contains 24,180 train questions and 6,046 test questions, and we adopt Wikidata queries as target labels
. Furthermore, we also conduct experiments on QALD-9 and QALD-10, both of which are more challenging KGQA datasets. 
We propose two evaluation metrics for measuring the performance of generation: Query Match (QM) and answer F1 score. QM determines accuracy by comparing the consistency between the predicted SPARQL query and the ground truth, while the answer F1 score evaluates the KG responses of the generated query against the expected answers, calculated as the harmonic mean of the precision and recall.

We employ T5 as the foundational model, and 
leverage T5-small, T5-base, and T5-large 
to explore the impact of model scale.
We use Adafactor as the optimizer and set the learning rate as 0.0001. The number of training epochs is 1024 during fine-tuning. 
All experiments are carried out on GeForce RTX 3090 GPUs, and it costs about 18 hours for additional pre-training on a single GPU.
\subsection{Main Results}
From Table \ref{tab:lc2_result}, we can observe that compared with the original T5 model, the triplet-order-sensitive pre-training effectively improves the performance, as our proposed TosT5 achieves state-of-the-art results across all model sizes on LC-QuAD 2.0 dataset. Specifically, for large-sized models, our method outperforms the baseline on both F1 score and QM metric, with an improvement of 1.3\% and 1.4\% respectively, reaching 95.4\% and 93.5\%, which establishes a new state-of-the-art performance. The performances of both the original T5 and our TosT5 model improve as the model size increases, which is consistent with the notion that larger pre-trained models are more capable of capturing intricate patterns and nuances in the text.

\begin{table}[!h]
\centering
\footnotesize
\caption{Experimental results for LC-QuAD 2.0. QM scores are not provided by \cite{banerjee2022modern} \cite{rony2022sgpt}. $^{\dag}$ denotes only gold entities are provided. $^{\star}$ denotes our re-implemented results. The values in parentheses indicate the 95\% confidence intervals. The best performance for each T5 size is bolded.}
\begin{tabular}{lll}
\toprule
Approach                         & F1    & QM      \\ 
\midrule
AQG-net \cite{chen2021formal}    & 44.9  & 37.4     \\
Multi-hop QGG \cite{lan2020query}  & 52.6  & 43.2    \\
CLC+BERT \cite{zou2021chinese}                    & 59.3  & 55.4   \\
BART \cite{banerjee2022modern}         & 64.0      & -            \\ 
PGN-BERT \cite{banerjee2022modern}     & 77.0    &  -           \\ 
PGN-BERT-BERT \cite{banerjee2022modern}& 86.0   & -         \\
SGPT$_{Q,K}$ \cite{rony2022sgpt}$^{\dag}$    & 89.0   & -  \\
T5-base  \cite{banerjee2022modern}     & 91.0        & -          \\ 
\midrule
T5-small$^{\star}$            & 92.4     &  90.2   \\
T5-base$^{\star}$          & 93.4     &  91.4   \\
T5-large$^{\star}$           & 94.1     &  92.1   \\
\midrule
TosT5-small       & \textbf{94.1}($\pm$0.22)     &  \textbf{92.0}($\pm$0.20)        \\
TosT5-base        & \textbf{95.0}($\pm$0.27)     &  \textbf{93.2}($\pm$0.24)        \\
TosT5-large       & \textbf{95.4}($\pm$0.21)     &  \textbf{93.5}($\pm$0.21)       \\
\midrule
T5-large$^{\dag}$         & 85.3     & 74.9    \\
TosT5-large$^{\dag}$       & 90.3($\pm0.31$)     &  79.2($\pm0.27$)       \\
\bottomrule
\end{tabular}
\label{tab:lc2_result}
\end{table}

Table \ref{tab:QALD_result} shows that our model exhibits greater improvements on the two challenging datasets, revealing our model's enhanced comprehension of SPARQL, which allows it to more effectively showcase its strengths in complex scenarios and achieve state-of-the-art results. Moreover, the performance improvement tends to gradually diminish as the model scale increases. This may be due to the fact that larger models are more capable of independently learning knowledge of triplets during training on downstream tasks, without relying entirely on our additional pre-training.

\begin{table}[t]
\centering
\footnotesize
\caption{Experimental results on QALD-9 and QALD-10 test set. The improvements compared to baselines are bolded.}
\begin{tabular}{lccccc}
\toprule
\multirow{2}{*}{Approach} & \multicolumn{2}{c}{QALD-9} & \multirow{2}{*}{} & \multicolumn{2}{c}{QALD-10} \\ \cmidrule{2-3} \cmidrule{5-6} 
                          & QM             & F1             &                   & QM           & F1           \\ \midrule
TeBaQA\cite{bfraunhofer2021knowledge}$^{\dag}$  & - & 28.8 & & - & - \\
SGPT$_{Q,K}$ \cite{rony2022sgpt}$^{\dag}$     & - & 67.8 & & - & - \\ \midrule

T5-small                  & 55.1          & 64.4          &                   & 33.9        & 38.8        \\
T5-base                   & 58.1          & 69.5          &                   & 36.4        & 39.8        \\
T5-large                  & 59.6          & 71.6          &                   & 35.7        & 45.3            \\ \midrule
TosT5-small               & 61.1\textbf{(+6.0)}          & 72.8\textbf{(+8.4)}          &                   & 40.1\textbf{(+6.2)}        & 47.2\textbf{(+8.4)}        \\
TosT5-base                & 61.8\textbf{(+3.7)}          & 75.8\textbf{(+6.3)}          &                   & 39.1\textbf{(+2.7)}        & 51.4\textbf{(+11.6)}            \\
TosT5-large               & 63.2\textbf{(+3.6)}         & 73.2\textbf{(+1.6)}          &                   & 39.0\textbf{(+3.3)}        & 51.2\textbf{(+5.9)}            \\ \bottomrule
\end{tabular}
\label{tab:QALD_result}
\end{table}

\subsection{Error Analysis}
\label{sec:errors}

\begin{figure}[htbp]
\subfigure[LC-QuAD 2.0]{
\begin{minipage}[t]{0.5\linewidth}
\includegraphics[width=4cm]{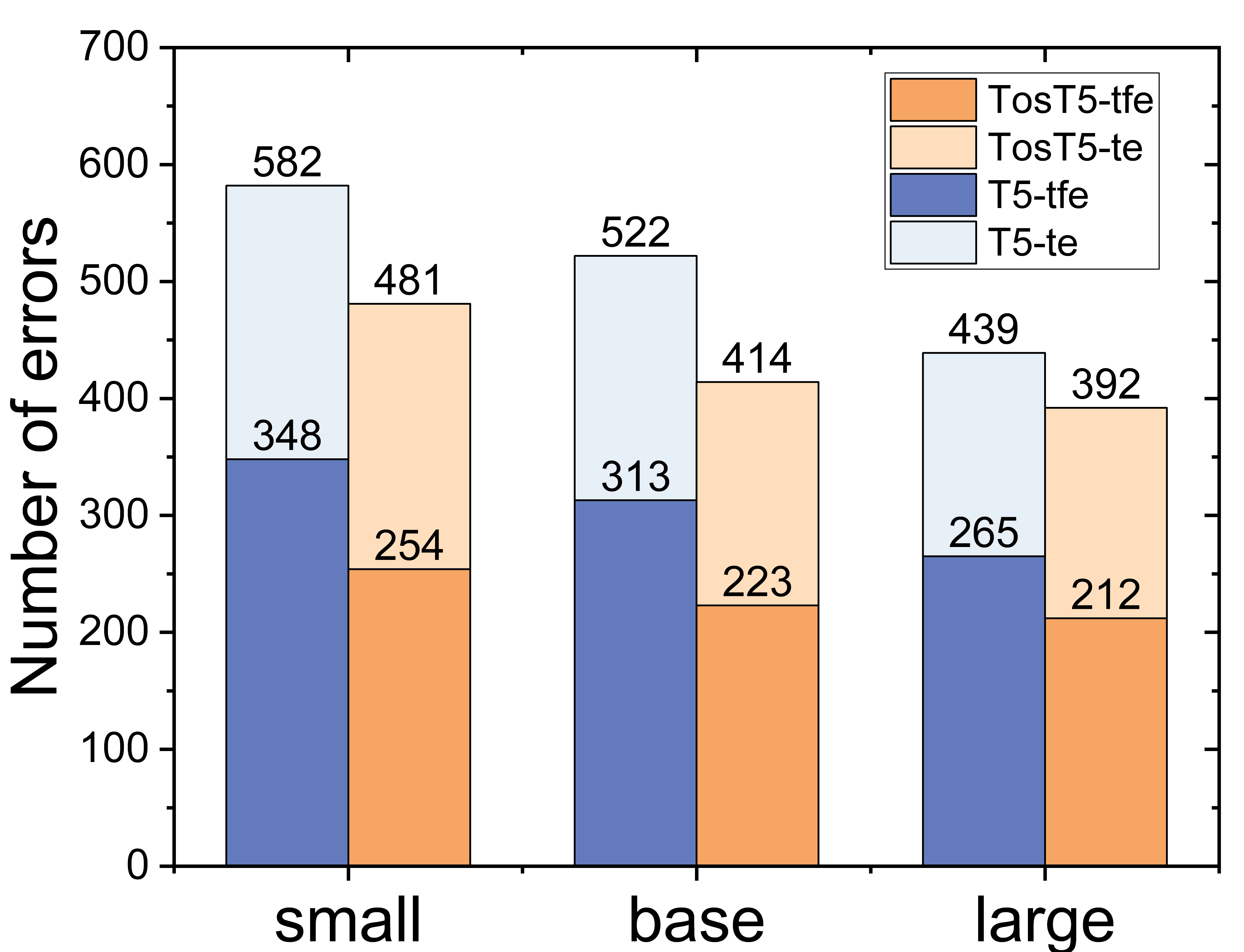}
\end{minipage}%
}%
\subfigure[QALD-9]{
\begin{minipage}[t]{0.5\linewidth}
\includegraphics[width=4cm]{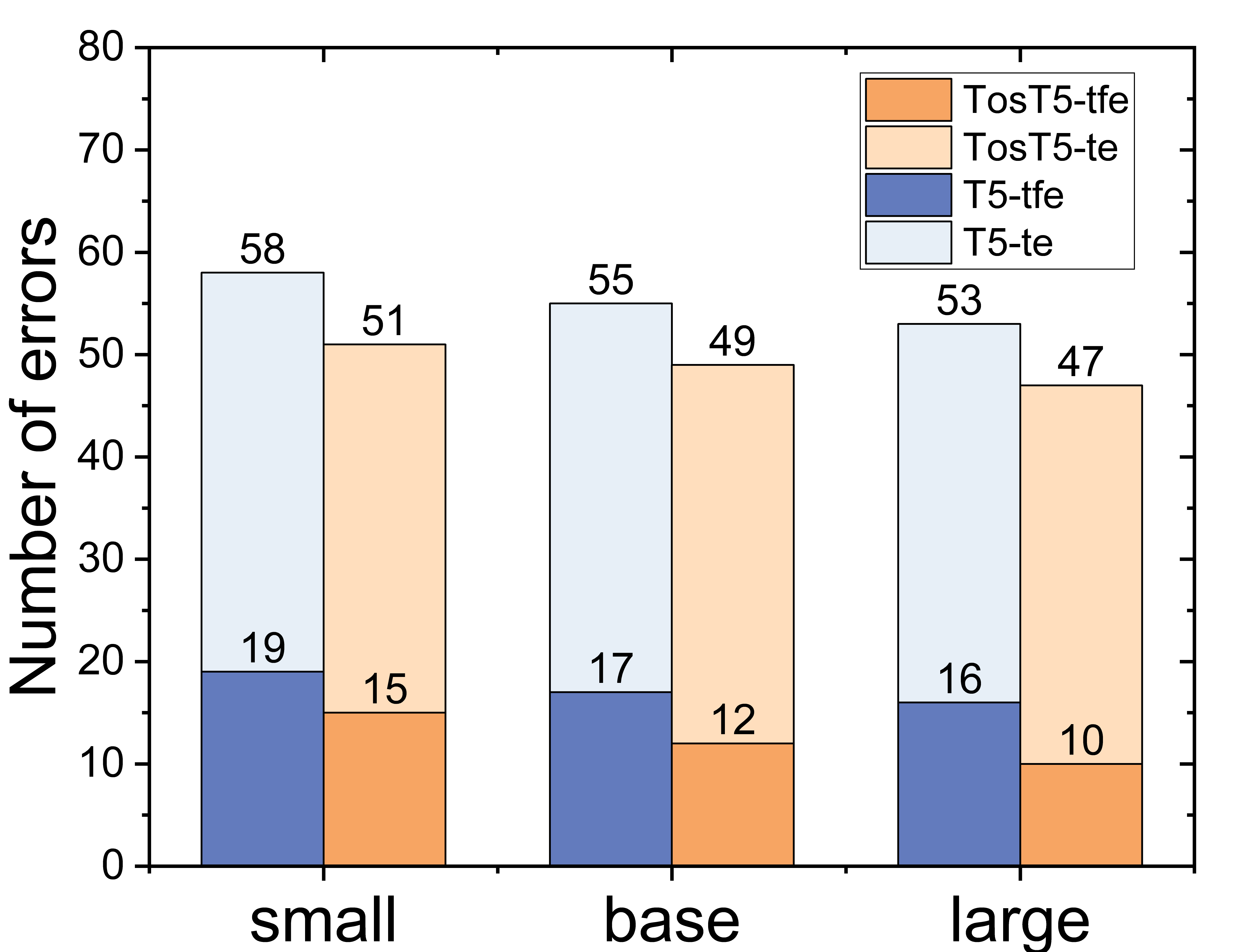}
\end{minipage}%
}%
\caption{Error analysis on models' prediction. "tfe" is short for "triplet-flip error", and "te" is short for "triplet error".}
\label{fig:error_analysis}
\end{figure}

We conduct a thorough error analysis to validate that our method effectively reduces triplet-flip errors, and the detailed results are shown as Figure \ref{fig:error_analysis}. Triplet errors contain triplet-flip errors and any other errors occurring within triplets, such as incorrect prediction for entities or relations. The results show that for the LC-QuAD 2.0 dataset, our approach reduces triplet errors by approximately 20\% across all model sizes, with a significant decrease of roughly 25\% in triplet-flip errors. For QALD-9, there is an approximate 15\% reduction in triplet errors and around 30\% decrease in triplet-flip errors. These results confirm that the pre-training helps to grasp more knowledge about triplets, especially their elements' order.

\subsection{Ablation Study}
We carry out a series of ablation experiments, and the results are shown in Table \ref{tab:ablation}. Firstly, verbalizing IRIs has a significant impact on performance, with a reduction of 1.91\% when it is omitted. Next, we investigate the effectiveness of the two pre-training objectives (\textit{i.e.}, TOC and MLM), and find that removing TOC results in a performance decrease of 0.98\% point, while omitting MLM leads to a performance decrease of 0.40\%. These results indicate that both the two objectives are efficient in our approach.

To further verify the effectiveness of the novel TOC we devised, we design two variants of TOC, namely Sentence Order Correction (SOC) and Triplet Flip Correction (TFC). SOC randomly shuffles a certain percentage of tokens in the SPARQL query and then lets the model correct the sequence. TFC merely reverses the positions of the subject entity and object entity in each triplet with a certain probability, challenging the model to reconstruct the original order. Their performances are inferior to our proposed TOC objective.

\begin{table}[!h]
\centering
\footnotesize
\caption{The ablation for verbalizing IRIs and pre-training objectives. TFC and SOC are different variants of TOC.}
\begin{tabular}{lcc}
\toprule
Approach    & QM    & Performance Drop \\ \midrule
TosT5       & 92.03 & -               \\
\quad w/o  Verbalizing IRIs    & 90.12     & \textbf{-1.91}            \\
\quad w/o   TOC   & 91.05 & \textbf{-0.98}            \\
\quad w/o   MLM   & 91.63 & \textbf{-0.40}            \\ \midrule
\quad TFC instead of TOC     & 91.61 & \textbf{-0.42}           \\
\quad TFC+MLM & 91.86 & \textbf{-0.17}            \\
\quad SOC instead of TOC     & 91.32 & \textbf{-0.71}            \\
\quad SOC+MLM & 91.50 & \textbf{-0.53}            \\ \bottomrule
\end{tabular}
\label{tab:ablation}
\end{table}

\subsection{Tough Scenario}
    Relation linking is not always a straightforward task, since the relations mentioned in questions sometimes differ from their labels in the KG\cite{dubey2018earl}. Thus, we introduce a tough scenario for our models where they are not provided with gold relations during training and inference. Instead, they learn in their parameters. We conduct experiments on LC-QuAD 2.0 dataset and the results are shown in Table \ref{tab:lc2_result}, marked by "$^{\dag}$". Our TosT5-large model achieves a QM of $79.2\%$ and an F1 score of $90.3\%$, outperforming both the T5-large baseline and the superior SGPT$_{Q,K}$ model. The results suggest that our pre-training objectives help the model learn the KG better, and our model is still powerful under this tough scenario.

\section{CONCLUSION}
In this paper, we introduce a triplet-order-sensitive T5 model, namely TosT5, specifically designed to address the triplet-flip errors exhibited in the Text-to-SPARQL task. 
We adopt T5 as the backbone and utilize two pre-training objectives to enhance the model's comprehension of the SPARQL language. 
We also propose to verbalize IRIs during training to better leverage the pre-trained model’s ability in language understanding. 
After undergoing the pre-training, our TosT5 model is fine-tuned on the downstream Text-to-SPARQL task, achieving new state-of-the-art performances on three well-known KGQA datasets. We believe our model will help enhance the overall performance of the KGQA system.

\begin{acks}
This work was sponsored by the National Key Research and Development Program of China (No. 2023ZD0121402) and the National Natural Science Foundation of China (NSFC) grant (No.62106143).
\end{acks}



\bibliographystyle{ACM-Reference-Format}
\balance
\bibliography{sample-base}


\end{document}